\documentclass[aps,pra,10pt,twocolumn,superscriptaddress,reprint]{revtex4-1}
\usepackage{amssymb}
\usepackage{graphicx}
\usepackage{bm}
\usepackage{hyperref}
\usepackage{mathtools}
\usepackage{bbold}
\usepackage{epsfig}
\usepackage{accents}

\newlength{\dhatheight}
\newcommand{\doublehat}[1]{%
    \settoheight{\dhatheight}{\ensuremath{\hat{#1}}}%
    \addtolength{\dhatheight}{-0.35ex}%
    \hat{\vphantom{\rule{1pt}{\dhatheight}}%
    \smash{\hat{#1}}}}

\newcommand{\pdiff}[2]{\frac{\partial #1}{\partial #2}}

\begin{document}
\def\one{{\mathchoice {\rm 1\mskip-4mu l} {\rm 1\mskip-4mu l} {\rm
\mskip-4.5mu l} {\rm 1\mskip-5mu l}}}

\title{Efficient continuous wave noise spectroscopy beyond weak coupling}
\date{\today}
\author{Kyle Willick}
\affiliation{Institute for Quantum Computing, University of Waterloo, Waterloo, Ontario, Canada N2L 3G1}
\affiliation{Waterloo Institute for Nanotechnology, University of Waterloo, Waterloo, Ontario, Canada N2L 3G1}
\affiliation{Department of Physics and Astronomy, University of Waterloo, Waterloo, Ontario, Canada N2L 3G1}
\author{Daniel K. Park}
\affiliation{Natural Science Research Institute, Korea Advanced Institute of Science and Technology, Daejeon, Republic of Korea 34141}
\author{Jonathan Baugh}
\email[Contact: ]{baugh@uwaterloo.ca}
\affiliation{Institute for Quantum Computing, University of Waterloo, Waterloo, Ontario, Canada N2L 3G1}
\affiliation{Waterloo Institute for Nanotechnology, University of Waterloo, Waterloo, Ontario, Canada N2L 3G1}
\affiliation{Department of Chemistry, University of Waterloo, Waterloo, Ontario, Canada N2L 3G1}

\begin{abstract}
The optimization of quantum control for physical qubits relies on accurate noise characterization. Probing the spectral density $S(\omega)$ of semi-classical phase noise using a spin interacting with a continuous-wave (CW) resonant excitation field has recently gained attention. CW noise spectroscopy protocols have been based on the generalized Bloch equations (GBE) or the filter function formalism, assuming weak coupling to a Markovian bath.
However, this standard protocol can substantially underestimate $S(\omega)$ at low frequencies when the CW pulse amplitude becomes comparable to $S(\omega)$. Here, we derive the coherence decay function more generally by extending it to higher orders in the noise strength and discarding the Markov approximation. Numerical simulations show that this provides a more accurate description of the spin dynamics compared to a simple exponential decay, especially on short timescales. Exploiting these results, we devise a protocol that uses an experiment at a single CW pulse amplitude to extend the spectral range over which $S(\omega)$ can be reliably determined to $\omega=0$.
\end{abstract}

\maketitle

\section{Introduction}
The problem of a qubit interacting with a noisy environment (bath) is of fundamental importance in the field of quantum information processing. Choosing the optimal strategy to fight decoherence depends on the noise characteristics of a particular qubit implementation. For many solid-state qubits, single-axis phase noise is dominant, and treating the environment in a stochastic semi-classical approximation suffices to describe the dephasing process. For example, a system-environment Hamiltonian of the form $H_{SE}=\sigma_z^{(s)}\sum_{i}\lambda_i \sigma_z^{(i)}/4$, where $\sigma^{(s)}$ ($\sigma^{(i)}$) is the Pauli operator for the system (environment) and $\lambda_i$ is the coupling strength, is approximated as the semi-classical $H_{sc}(t) = f(t)\sigma_z/2$ by tracing over the environmental degrees of freedom \cite{suterPRA}. In the limit of many environment qubits forming a spin bath, with intra-bath couplings strong compared to $\lambda_i$, $f(t)$ can be treated as a stationary, Gaussian-distributed function with zero mean, i.e. $\langle f(t)\rangle=0$. These properties will be assumed throughout the remainder of the paper.

In the context outlined above, knowledge of the spectral density function $S(\omega)$, the Fourier transform of the two-point correlation function for $f(t)$, can be used to optimize quantum control, such as dynamical decoupling (DD) and dynamically corrected gates~\cite{PhysRevLett.102.080501}. 
The spectral information can also be used in decoherence suppression techniques such as hole-burning \cite{revRef3.PhysRevLett.115.033601}.
 One way to estimate $S(\omega)$ is to monitor the response of the qubit as it undergoes DD pulse sequences with certain spectral properties \cite{PRL.107.170504,suterPRL,AlmogJPhysB,N.Bar-Gill_nature,PhysRevB.72.134519,bylander,fei_yan_nature,PhysRevLett.109.153601,0953-8984-29-33-333001}.
This can be understood intuitively using the overlap integral approach \cite{PhysRevLett.109.020501,Biercuk_NJP_2013,PhysRevLett.93.130406,PhysRevLett.87.270405,Cywinski,suterPRA,JPB_Biercuk,uhrigNJP,PhysRevLett.98.100504,PhysRevLett.113.250501,0953-8984-29-33-333001}. For example, under a series of equally spaced, instantaneous $\pi$ pulses, the bath-traced Hamiltonian becomes  $H_{sc}(t) = y(t)f(t)\sigma_z/2$ where $y(t)$ alternates between $+1$ and $-1$ at a period corresponding to the pulse spacing $\tau$, and the frequency of the decoupling cycle is $\Omega=\pi/\tau$. In this case, an exponential decay of qubit coherence is predicted, $\langle \sigma_x(T) \rangle = \langle \sigma_x(0) \rangle e^{-\chi(T)}$, where the time-dependent decay rate is determined by the overlap integral of the noise spectral density and the frequency-domain filter function (the Fourier transform of the time-domain filter $y(t)$), $|F(\omega,T)|^2$:
\begin{equation}
\label{eq:relaxation}
\chi(T)=\int^\infty_{-\infty} d\omega S(\omega)|F(\omega,T)|^2.
\end{equation}
In other words, the external control sequence acts as a bandpass filter and can be tailored such that the qubit is most sensitive to certain spectral bands of the noise power. The same formalism can be applied in quantum sensing such as spin-based magnetometry for oscillating fields using nitrogen-vacancy centers in diamond~\cite{NaturePhysics.4.810.2008,PhysRevLett.106.080802,PhysRevA.86.062320,Staudacher561,NatureNano.10.129.2015,NatureNano.10.125.2015}, further motivating the development of accurate DD-based spectroscopy over an extended bandwidth. If the function $|F(\omega,T)|^2$ is spectrally broad or peaked at many frequencies, extracting $S(\omega)$ becomes challenging, particularly if the functional form of $S(\omega)$ is not known a priori. $|F(\omega,T)|^2$ can be made spectrally narrow if there are sufficiently many decoupling cycles, i.e., $T>>2\tau$ where $2\tau$ is the decoupling period. The filter function then approaches a delta-function at the decoupling cycle frequency and its harmonics $k\Omega$, where $k$ is a nonzero integer. When $T$ is much longer than the time scale of the bath correlation decay, $S(\omega)$ can be regarded as constant within the probed spectral width, allowing equation~\ref{eq:relaxation} to be written in a discrete form. A protocol for estimating $S(\omega)$ using the discrete form of equation~\ref{eq:relaxation} and by taking the harmonics into account was designed and implemented experimentally in Ref.~\cite{suterPRL}.

The pulsed method becomes disadvantageous at high probe frequencies such that finite pulse width effects cannot be ignored and limit the minimum pulse spacing. Moreover, the lowest frequency (given by the maximum pulse delay) dictates the frequency resolution with which the spectral density function is probed~\cite{suterPRL}. This makes the protocol inefficient in certain situations. For example, probing an unknown noise spectrum over a wide frequency window requires a very large number of experiments.

An alternative approach is to monitor coherence decay under a continuous wave (CW) ``spin-locking" pulse, also known in the NMR literature as a $T_{1\rho}$ measurement. In this case, the qubit dynamics can only be studied perturbatively due to the non-commutativity of the effective Hamiltonian. $T_{1\rho}$ experiments have been used in NMR to probe slow atomic motions that give rise to fluctuations in the dipolar field \cite{PhysRev.135.A1099,PhysRev.137.A235,ailionAMR,LookAndLowe}. 
The NMR literature, however, has not directly addressed the problem of extracting an unknown and arbitrary $S(\omega)$ from a series of $T_{1\rho}$ measurements. This was first addressed in the context of the generalized Bloch equations (GBE) formalism \cite{GBE,PhysRevB.72.134519}. The generalized Bloch equations were derived to describe the relaxation dynamics of a system simultaneously interacting with a heat bath and an arbitrarily strong excitation field. The derivation is based on the following assumptions: (1) the system and the bath are weakly coupled, and are initially in a product state; (2) the time scale of the relaxation of the system is much slower than that associated with the decay of the bath correlation functions and the period of the driving field; 
(3) the bath-induced coherent system dynamics are negligible compared to that induced by the system Hamiltonian; (4) the rotating wave approximation (RWA). The weak coupling assumption means that we keep terms only up to second order in the system-bath coupling strength $f(t)$. The second assumption leads to the aforementioned delta-function approximation. For the noise model described earlier, the GBE predicts an exponential decay of coherence $\langle \sigma_x \rangle$ in the $T_{1\rho}$ experiment. The decay rate is directly proportional to $S(\Omega)$ where $\Omega$ is the pulse amplitude (Rabi frequency). The steady-state coherence is negligible as long as the high temperature limit $k_B \text{T} >> \hbar \Omega$ is satisfied. Note that the decay rate here is time independent and thus cannot capture non-Markovian dynamics. The CW approach can often perform well to higher frequencies than the pulsed method, since finite pulse width effects tend to appear before the maximum excitation power is reached or before the RWA is violated. Moreover, the CW protocol can be more efficient than pulsed methods since a single coherence decay measurement yields the spectral density of noise at the target frequency. 
$T_{1\rho}$ noise spectroscopy was demonstrated experimentally in Ref.~\cite{AlmogJPhysB} for optically-trapped ultracold atoms coupled to a collisional bath, and in Ref.~\cite{fei_yan_nature} in the context of superconducting qubit decoherence. In the latter case, the analysis was based on the GBE but included more general noise (relaxation) than considered here.

Neither the CW or pulsed methods can probe to arbitrarily low frequencies using the standard analyses above. In these analyses, the number of drive field periods (decoupling cycles) should be large enough to justify the delta function approximation, i. e. $\Omega T/(2\pi)>>1$. Since the signal decay timescale is $T\sim 1/S(\Omega)$, the minimum probe frequency is limited by the condition $\Omega >> 2\pi S(\Omega)$ (where $\Omega = \pi/\tau$ in the pulsed method). The main goal of this paper is to study spin dynamics under CW excitation beyond approximations (1) and (2) above, so that the signal decay at low frequencies $\Omega \sim 2\pi S(\Omega)$ can be better modeled. This information is then used to increase the spectral range over which noise spectroscopy produces valid results. We describe the state evolution in the Liouville representation \cite{liouville} and apply the cumulant expansion method \cite{cumulant1,cumulant2} to calculate the ensemble average, finding the functional form of the coherence decay up to fourth order in $f(t)$ (or second order in $S(\omega)$). The resulting equations are derived without any assumptions about the CW pulse length or the bath correlation time, in order to capture non-Markovian behaviour. These results are used to design a CW noise spectroscopy protocol that extends the range for which $S(\omega)$ can be accurately determined down to $\omega = 0$.

The remainder of the paper is organized as follows. Section~\ref{sec:2} derives the coherence decay function in the $T_{1\rho}$ experiment up to fourth order in $f(t)$ (i.e., second order in $S(\omega)$). Section~\ref{sec:3} compares our results to the standard exponential decay function and shows that our model predicts the signal decay significantly better in the short time regime. Section~\ref{sec:4.1} presents the noise spectroscopy protocol exploiting the coherence decay function derived in Sec.~\ref{sec:2}, and the improved accuracy in the $S(\omega)$ determination is demonstrated in Sec.~\ref{sec:4.2} via numerical simulations. Section~\ref{sec:5} concludes.

\section{Coherence decay function}
\label{sec:2}
In this section we derive the coherence decay function of a system under CW driving as a function of the spectral density, $S(\omega)$, of Gaussian, zero-mean semi-classical phase noise as introduced above. In the interaction frame of the CW pulse of amplitude $\Omega$ along $\sigma_x$ in the rotating frame, the semi-classical stochastic Hamiltonian transforms in time $t$ as
\begin{equation}
\label{eq:intHam}
\tilde{H}_{sc}(t)=f(t)(\cos(\Omega t)\sigma_z + \sin(\Omega t)\sigma_y)/2.
\end{equation}
The derivation of the coherence decay function under this Hamiltonian must involve perturbation series, since $[\tilde{H}_{sc}(t_1),\tilde{H}_{sc}(t_2)]\neq 0]$. Using stochastic Liouville theory and super operator formalism \cite{liouville}, the ensemble averaged qubit evolution can be described as $\langle\doublehat{\rho}(T)\rangle=\langle\Lambda(T)\rangle\doublehat{\rho}(0)$, where $\rho(T)$ is the density matrix decribing the qubit at time $T$, $\langle\cdot\rangle$ denotes ensembles averaging over noise realizations, $\doublehat{\cdot}$ denotes vectorization that stacks the rows of a $d\times d$ matrix into a $d^2 \times 1$ vector, and 
\begin{multline}
\label{eq:Lambda}
\Lambda(T) = \mathcal{T}\exp\left\lbrace -i\int^{T}_0 \mathcal{L}(t) dt\right\rbrace, \\
\mathcal{L}(t)=\tilde{H}^*_{sc}(t)\otimes\one-\one\otimes \tilde{H}_{sc}(t),
\end{multline}
where $\one$ is the unit matrix, and $\mathcal{T}$ is the Dyson time ordering operator \cite{cumulant2}. The ensemble average of the noisy operator $\Lambda(T)$ can be evaluated with the cumulant expansion
\begin{equation}
\label{eq:cumdef}
\langle\Lambda(T)\rangle=\exp\left\lbrace K(T) \right\rbrace,\quad K(T)=\sum_{n=1}^{\infty}\frac{(-iT)^n}{n!}k_n,
\end{equation}
where $K(T)$ is called the cumulant function and $k_n$ is called the $n$th cumulant~\cite{cumulant1,cumulant2}. By Taylor-expanding and comparing equations~\ref{eq:Lambda} and \ref{eq:cumdef}, the cumulants can be found. 

For the Hamiltonian in equation~\ref{eq:intHam}, the powers of $\mathcal{L}(t)$ are linear combinations of the operators from following set:
\begin{multline}\label{eq:opSet}
\mathcal{N}=\lbrace\one\otimes\one,\; \sigma_x\otimes\sigma_x,\; \sigma_y\otimes\sigma_y,\; \sigma_z\otimes\sigma_z,\; \\  \sigma_x\otimes\one-\one \otimes \sigma_x,\; \sigma_y\otimes\sigma_z-\sigma_z\otimes\sigma_y\rbrace,
\end{multline}
Moreover, $k_{n}$ is proportional to the $n$th power of $\mathcal{L}(t)$. As a result, $k_{n}$ is a linear combination of the operators in $\mathcal{N}$. Thus, $\langle \Lambda(T) \rangle$ can be expressed as
\begin{equation}
\label{eq:LambdaAve}
\langle\Lambda(T)\rangle=\exp\left(\sum_{m=1}^{|\mathcal{N}|}a_m\mathcal{N}_m \right),
\end{equation}
where $\mathcal{N}_m$ is the $m$th element in $\mathcal{N}$.
In the spin-locking experiment, the normalized signal is
\begin{align}
\label{eq:aveSig}
\langle \sigma_x(T)\rangle &= \frac{\doublehat{\sigma}_x^{\text{T}}\langle \Lambda(T)\rangle \doublehat{\sigma}_x}{\doublehat{\sigma}_x^{\text{T}} \doublehat{\sigma}_x}\\ &= \resizebox{.8\hsize}{!}{$\frac{\langle \Lambda(T)\rangle_{2,2}+\langle \Lambda(T)\rangle_{2,3}+\langle \Lambda(T)\rangle_{3,2}+\langle \Lambda(T)\rangle_{3,3}}{2}$}
\end{align}
where $\langle \cdot \rangle _{i,j}$ is the element of an operator at row $i$ and column $j$.  Combining equations \ref{eq:LambdaAve} and \ref{eq:aveSig},
\begin{equation}
\label{eq:SigGeneral1}
\langle \sigma_x(T)\rangle=\exp\left( a_1+a_2+a_3-a_4\right).
\end{equation}
It is convenient to write above equation as
\begin{align}
\langle \sigma_x(T)\rangle &= \exp\left(\sum_n a_{1,n}+a_{2,n}+a_{3,n}-a_{4,n} \right) \\ &=\exp\left(\sum_n\chi_n(T)\right),
\label{eq:SigGeneral2}
\end{align}
where $\sum_na_{m,n}=a_m$, and the index $n$ indicates that the contribution is linked to the $n$th cumulant. The above equation has several significant implications. First, the average signal in the $T_{1\rho}$ measurement is an exponential function whose argument (decay rate) is given as a perturbation series. Second, the $n$th order decay rate is proportional to $k_{n}$, and therefore proportional to the ensemble average of products of Liouvillians,
$\langle\prod_{j=1}^{n}\mathcal{L}(t_j)\rangle$, and hence proportional to the average over products of the classical Gaussian distributed random variable, $\langle\prod_{j=1}^{n}f(t_j)\rangle$). Third, the $n$th order decay rate can simply be calculated from the coefficients of the $\one\otimes\one$, $\sigma_x\otimes\sigma_x$, $\sigma_y\otimes\sigma_y$, and $\sigma_z\otimes\sigma_z$ terms corresponding to the $n$th order cumulant, independently from other cumulants. Moreover, since $f(t)$ is Gaussian-distributed with zero mean, for an integer $n$, $\langle f(t_1)\dots f(t_{2n-1})\rangle=0$ according to Isserlis’ Gaussian moment theorem \cite{10.2307/2331932}. Subsequently, only even order cumulants are non-zero, and the first few terms are
\begin{widetext}
\begin{align}
\label{eq:k2}
&k_2=\frac{1}{T^2}\mathcal{T}\int^{T}_0dt_1\int^{T}_0dt_2\langle\mathcal{L}(t_1)\mathcal{L}(t_2)\rangle,\\
\label{eq:k4}
&k_4+3k_2^2=\frac{1}{T^4}\mathcal{T}\int^T_0dt_1\int^T_0dt_2\int^T_0dt_3\int^T_0dt_4\langle\mathcal{L}(t_1)\mathcal{L}(t_2)\mathcal{L}(t_3)\mathcal{L}(t_4)\rangle,\\
\label{eq:k6}
&k_6+15(k_2k_4+k_2^3)=\frac{1}{T^6}\mathcal{T}\int^T_0dt_1\cdots\int^T_0dt_6\langle\mathcal{L}(t_1)\cdots\mathcal{L}(t_6)\rangle.
\end{align}
\end{widetext}

In the following, we present analytical calculation of the first two non-zero decay terms. We also show that in the limit of $T\rightarrow \infty$, the $2$nd order decay rate is identical to the GBE result. To go beyond the GBE-based analysis, we omit the large $T$ approximation, and describe the qubit dynamics including the $4$th order decay rate.

\subsection{Second-order decay rate}
\label{sec:2.1}
By expanding equation~\ref{eq:k2} (the $2$nd order cumulant) using the operators in equation~\ref{eq:opSet}, we find $a_{m,2}$. Then from equation~\ref{eq:SigGeneral2} we acquire the decay rate attributed to the $2$nd order cumulant
\begin{align}
\chi_2(T) &=-\int^{T}_0dt_1\int^{t_1}_0 dt_2\langle f(t_1)f(t_2)\rangle \cos\left(\Omega\left(t_1-t_2\right)\right) \\
\label{eq:chi2}
&=-\frac{1}{2\pi} \int^{\infty}_{-\infty}d\omega S(\omega)\mathcal{F}_2(\omega,\Omega,T)
\end{align}
where \begin{equation*} \mathcal{F}_2(\omega,\Omega,T) = \int^{T}_0dt_1\int^{t_1}_0dt_2 \text{e}^{i\omega\left(t_1-t_2\right)}\cos\left(\Omega\left(t_1-t_2\right)\right).
\end{equation*} 
The imaginary part of $\mathcal{F}_2(\omega,\Omega,T)$ is an odd function with respect to $\omega$, and $S(\omega)$ is an even function, thus equation~\ref{eq:chi2} can be further simplified as
\begin{equation}
\label{eq:chi2R}
\chi_2(T)=-\frac{1}{2\pi} \int^{\infty}_{-\infty}d\omega S(\omega)\text{Re}\left(\mathcal{F}_2\left(\omega,\Omega,T\right)\right).
\end{equation}

\begin{figure}
\centering
\includegraphics[width=1\columnwidth]{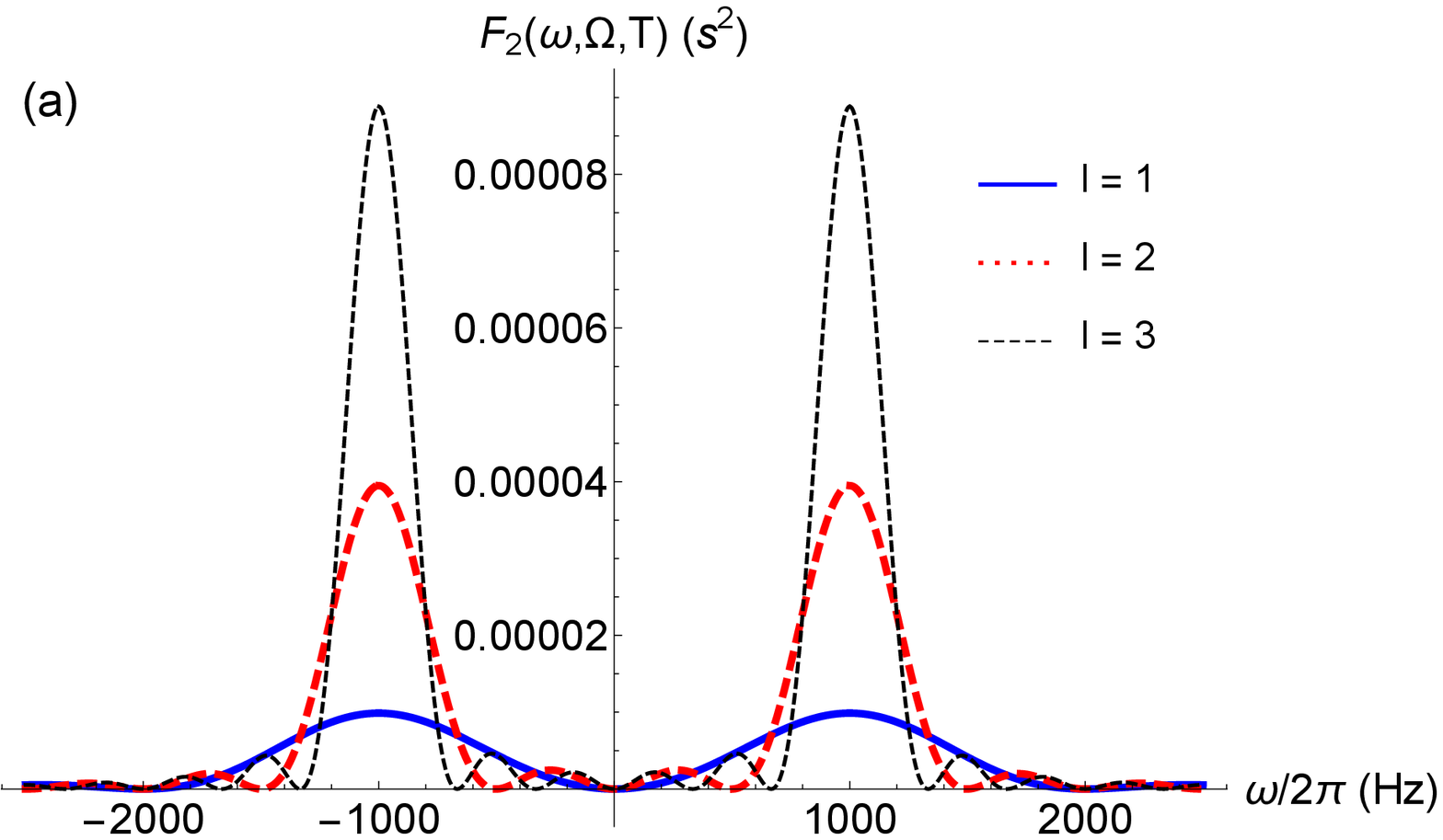}\vspace{2mm}
\includegraphics[width=1\columnwidth]{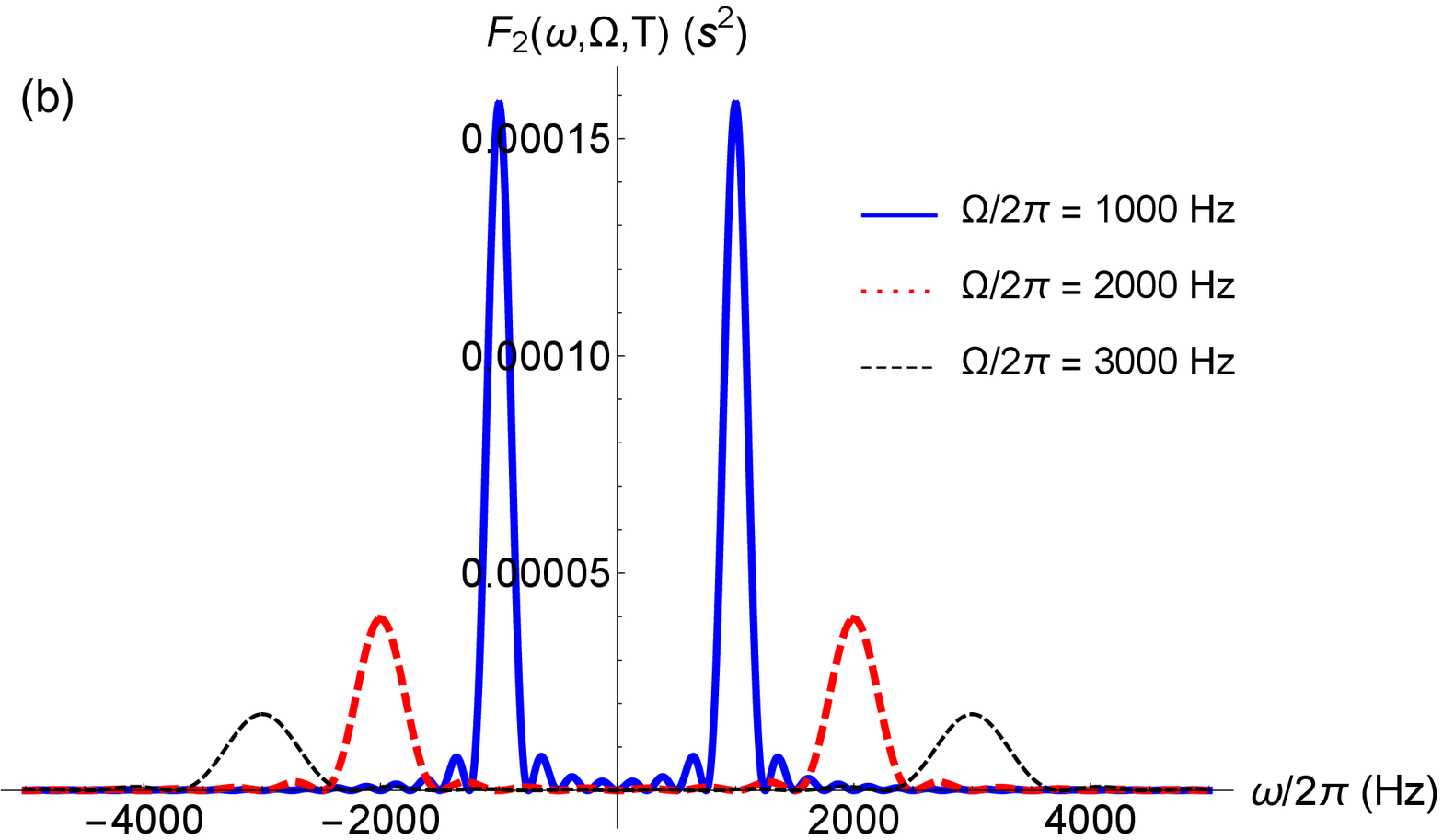}
\caption{\label{fig:F2}The real parts of $\mathcal{F}_2(\omega,\Omega,T=2\pi l/\Omega)$, where $l$ is the number of Rabi cycles, for (a) $\Omega/2\pi=1000$ Hz and $l=1,\;2,\;3$, and (b) for $l=4$ and $\Omega/2\pi=1000,\;2000,\;3000$ Hz.}
\end{figure}

Figure~\ref{fig:F2} shows the real parts of $\mathcal{F}_2(\omega,\Omega,T)$ for fixed values of (a) $\Omega$ and (b) the number of Rabi cycles $l = \Omega T / 2 \pi$. The even function $\text{Re}\left(\mathcal{F}_2\left(\omega_,\Omega,T\right)\right)$ behaves like $\delta(\omega  \pm \Omega)$ as $l\rightarrow \infty$ (i.e., as $T\rightarrow \infty$). Thus, for a fixed value of $\Omega$ and in the limit of $T\gg 1$, the $2$nd order decay rate can be approximated as
\begin{align}
\label{eq:chi2gbe}
\chi_2(T)&\approx -\frac{S(\Omega)}{\pi}\int^{\infty}_{0}d\omega\text{Re}\left(\mathcal{F}_2\left(\omega_,\Omega,T\right)\right)\nonumber\\
&=- S(\Omega)T/2.
\end{align}
Therefore, the normalized spin signal attributed to the $2$nd order cumulant  in the limit of large $T$ is a simple exponential decay:
\begin{equation}
\label{sigLargeN}
\langle\sigma_x(T)\rangle\approx\exp\left\lbrace -S(\Omega)T/2\right\rbrace.
\end{equation}
This expression is equal to the result derived from the GBE in the high temperature limit $k_B \text{T} >> \hbar \Omega$.

\subsection{Fourth-order decay rate}
\label{sec:2.2}
Following the same steps as for the $2$nd order term evaluation, the $4$th order cumulant $k_4$ can be expressed in terms of the operators in equation \ref{eq:opSet}, and the coefficients $a_{m,4}$ can be found. The decay rate attributed to the $4$th order cumulant can be expressed as
\begin{widetext}
\begin{equation}
\label{eq:chi4}
\chi_4(T)=\int^T_0dt_1\int^{t_1}_0dt_2\int^{t_2}_0dt_3\int^{t_3}_0dt_4\left\langle \prod_{i=1}^4f(t_i)\right\rangle\prod_{j=1}^2\cos\left(\Omega\left(t_{2j-1}-t_{2j}\right)\right) -\frac{1}{2}\left(\chi_2(T)\right)^2.
\end{equation}
\end{widetext}
We can use Isserlis Gaussian moment theorem again, and write the $4$th order correlation function as the products of the $2$nd order correlation functions: $\left\langle f(t_1)f(t_2)f(t_3)f(t_4)\right\rangle=\langle  f(t_1)f(t_2)\rangle\langle f(t_3)f(t_4)\rangle+\langle  f(t_1)f(t_3)\rangle\langle f(t_2)f(t_4)\rangle
+\langle  f(t_1)f(t_4)\rangle\langle f(t_2)f(t_3)\rangle.$
The product of the $2$nd order correlation function can be expressed in terms of the spectral density, $S(\omega)$ as before. Then, the $4$th order decay rate can be rewritten compactly as
\begin{widetext}
\begin{equation}
\label{eq:chi4_2}
\chi_4(T)=\left(\frac{1}{2\pi}\right)^2\int^{\infty}_{-\infty}d\omega_1\int^{\infty}_{-\infty}d\omega_2S(\omega_1)S(\omega_2)\mathcal{F}_4(\omega_1,\omega_2,\Omega,T),
\end{equation}
where
\begin{align}
&\mathcal{F}_4(\omega_1,\omega_2,\Omega,T)=\int^T_0dt_1\int^{t_1}_0dt_2\int^{t_2}_0dt_3\int^{t_3}_0dt_4\prod_{j=2}^1\cos\left(\Omega\left(t_{2j-1}-t_{2j}\right)\right))\nonumber\\
&\quad\times\left(e^{i\omega_1(t_1-t_2)}e^{i\omega_2(t_3-t_4)}+e^{i\omega_1(t_1-t_3)}e^{i\omega_2(t_2-t_4)}+e^{i\omega_1(t_1-t_4)}e^{i\omega_2(t_2-t_3)}\right)\nonumber\\
&-\frac{1}{2}\text{Re}\left(\mathcal{F}_2\left(\omega_1,\Omega,T\right)\right)\text{Re}\left(\mathcal{F}_2\left(\omega_2,\Omega,T\right)\right).
\end{align}
\end{widetext}
The above time integration can be carried out analytically. Symmetries in the filter function $\mathcal{F}_4$ result in the imaginary component of the integral going to zero as was the case for 2nd order decay rate.
Finally, using the symmetry of $\text{Re}(\mathcal{F}_4)$ the expression for the $4$th order decay rate can be  expressed as
\begin{small}
\begin{equation}
\label{eq:G4}
\chi_4(T)=\frac{1}{2\pi^2}\int^{\infty}_{0}d\omega_1\int^{\infty}_{0}d\omega_2S(\omega_1)S(\omega_2)\tilde{\mathcal{F}}_4(\omega_1,\omega_2,\Omega,T),
\end{equation}
\end{small}
where 
\begin{small}
\begin{equation*}
\tilde{\mathcal{F}}_4(\omega_1,\omega_2,\Omega,T)=\text{Re}\left(\mathcal{F}_4(\omega_1,\omega_2,\Omega,T)+\mathcal{F}_4(\omega_1,-\omega_2,\Omega,T)\right).
\end{equation*}
\end{small}

\section{Accuracy of coherence decay}
\label{sec:3}

To evaluate the accuracy of the cumulant expansion decays, we simulate a set of experiments with known input noise spectra, $S_{input}(\omega)$. The simulated signal decays are generated by time-discretized unitary evolution of an initial state density matrix, using $N = 10,000$ randomly generated noise realizations. A cosine series representation, as described in reference \cite{Shinozuka1991}, is used to generate the noise realizations, $f(t)$ in equation \ref{eq:intHam}. This simulation method accurately represents stationary, Gaussian noise matching the input noise spectrum, and converges to the correct spectrum at a rate of 1/N, where N is number of noise samples used. For all noise spectra in this paper, $S_{input}(\omega)$ plateaus below $\omega = 1\:\text{rad/s}$, i.e. $S_{input}(\omega') = S_{input}(1\:\text{rad/s})$ for $\omega' < 1\:\text{rad/s}$. These simulated signal decays are used in the following sections to represent experimental data. Since the number of noise realizations $N$ is finite, we recalculate $S_{in}(\omega)$ based on the simulated signal decays, and this is the $S_{in}(\omega)$ that appears in the plots below.  

\begin{figure}
\centering
\includegraphics[width=0.9\linewidth]{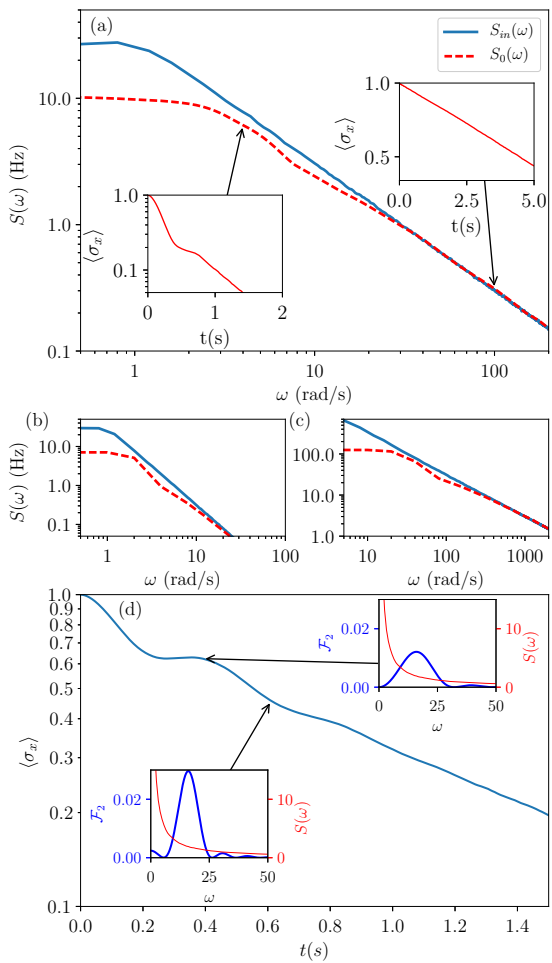}
\caption{\label{fig:GBEBreak} \textbf{a} Determination of $S(\omega)$ using simple exponential fitting in simulated CW noise spectroscopy experiments. The signal decay is simulated using an input $S_{in}(\omega) \approx 30\:\text{Hz}^2/\omega$, and least-squares fitting is used to calculate the noise spectrum $S_{0}(\omega)$. At small pulse amplitudes (low Rabi frequencies), the exponential fitting returns inaccurate spectral information. The insets show the signal decays at the indicated pulse amplitudes, illustrating the non-exponential signal decay behaviour at low $\Omega$. \textbf{b,c} Examples of the same procedure applied to different noise spectra, with b) $S_{in}(\omega) \approx 30\:\text{Hz}^3/\omega^2$, and c) $S_{in}(\omega) \approx 3000\:\text{Hz}^2/\omega$. \textbf{d} A signal decay taken at $\Omega = 16\:\text{rad/s}$ from the simulated experiment used in (a), illustrating non-exponential decay. The insets show $\mathcal{F}_2(\omega,\Omega,T)$ and $S(\omega)$ at two different time points, highlighting the importance of the low frequency component of $\mathcal{F}_2(\omega,\Omega,T)$ which overlaps with a large spectral noise density.}
\end{figure}

Using the simple exponential expression from equation \ref{sigLargeN}, $S(\omega)$ can be accurately determined when the assumptions listed in section 1 are valid, i.e. when the relaxation timescale is long compared to the drive field period ($\Omega/2\pi \gg S(\Omega)$). When this condition is not met, the signal decay can be non-exponential and the standard analysis can give inaccurate values for $S(\omega)$. Figure \ref{fig:GBEBreak}a shows the result of applying a least-squares exponential fit and using equation \ref{sigLargeN} to determine $S(\omega)$ for a $1/f$ input noise spectrum. In this example, $S_{input}(\omega) = 30\:\text{Hz}^2/\omega$ and $S_0(\omega)$ was obtained by fitting simulated CW noise spectroscopy experiments at $\omega = {1, 2, 4, 8, 12, 16, 20, 24, 32, 40, 64, 100, 125}\:\text{rad/s}$. The insets show examples of the normalized signal decay $\langle\sigma_x(t)\rangle$, which becomes non-exponential at low $\Omega$, leading to inaccurate $S(\Omega)$ values. Figures \ref{fig:GBEBreak}b and c show this same procedure applied to (b) a $1/f^2$ noise spectrum (c) and a $1/f$ noise spectrum with 100 times higher noise power than that shown in (a). Simulated CW experiments at $\omega = {1, 2, 4, 6, 8, 10, 12, 15, 20, 25, 30, 35, 40, 50, 120}\:\text{rad/s}$ were used for figure \ref{fig:GBEBreak}b, and experiments at $\omega = {1, 10, 20, 40, 80, 120, 160, 320, 640, 1000, 1250}\:\text{rad/s}$ were used for figure \ref{fig:GBEBreak}c. 

The non-exponential signal decay displayed in the inset of figure \ref{fig:GBEBreak}a can be understood based on the shape of the filter function with respect to $S(\omega)$ when the $\delta$-function approximation no longer holds. Figure \ref{fig:GBEBreak}d shows a non-exponential signal decay, with $\chi_2$ displayed in insets at certain time points in the decay, along with the noise spectrum $S(\omega)$. The finite width of the main filter function peak, as well as its satellite peaks, overlaps with large values of $S(\omega)$ at low frequencies, producing oscillations in the signal decay. 
 
\begin{figure}
\centering
\includegraphics[width = 1\linewidth]{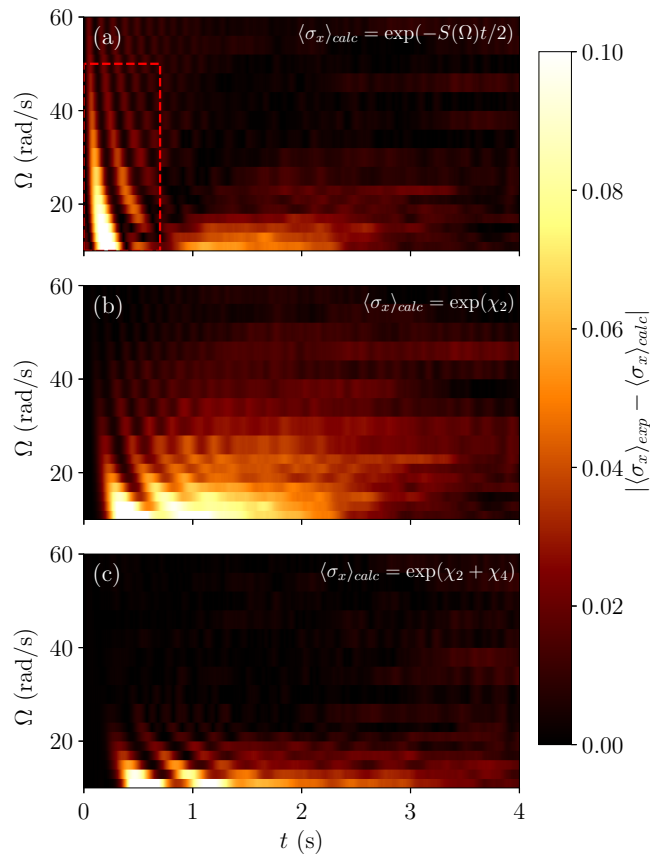}
\caption{\label{fig:GBEErrors} The error between the simulated ``experimental" signal decays and theoretical decays calculated by standard (i.e. equation \ref{sigLargeN}) and cumulant expansion methods. \textbf{a} Using the exponential form $\exp(-S(\Omega)t/2)$ results in large errors at short time scales ($t < 0.5\:\text{s}$), i.e. the region indicated by the dashed box. \textbf{b} and \textbf{c}, using the cumulant expansion method reduces the error. By using $\chi_2 + \chi_4$, the experimental decays are better matched at all times, down to low drive amplitudes ($\Omega \approx 20\:\text{rad/s}$ for the parameters chosen here).}
\end{figure}

Figure \ref{fig:GBEErrors} shows the error, as a function of decay time and drive field amplitude, between the simulated ``experimental'' signal decays and theoretical decays calculated using one of three different methods. Each calculation uses the same $S_{in}(\omega)$ that generated the simulated decays. Figure \ref{fig:GBEErrors}a shows the result of applying the simple exponential from equation \ref{sigLargeN}. Figures \ref{fig:GBEErrors}b and c show the decays calculated via the cumulant expansion from equation \ref{eq:SigGeneral2}, with (b) using only $\chi_2$, and (c) using both $\chi_2$ and $\chi_4$.  

The cumulant expansion method provides much better matching to signal decays at short times (t $<$ 0.5 s in the example in figure \ref{fig:GBEErrors}) compared to the standard exponential decay of equation \ref{sigLargeN}. However, using only $\chi_2$ introduce errors in the intermediate (20-35 rad/s) $\Omega$ region. Using the cumulant expansion up to fourth order ($\chi_2 + \chi_4$), the mismatch is reduced at all times over a large portion of the drive amplitudes ($\Omega \gtrsim 20\:\text{rad/s}$ for the parameters chosen in this example). 

\section{Noise spectroscopy based on the cumulant expansion}
\label{sec:4}
The $\chi_2+\chi_4$ cumulant expansion method can be used to improve CW noise spectroscopy when the experimental signal decays become non-exponential. The accuracy of a given noise spectrum estimate $S'(\omega)$ can be tested by comparing the cumulant expansion signal decay, calculated using $S'(\omega)$, and the experimental decay. Furthermore, the non-exponential, oscillatory behaviour observed at short timescales is the result of a wide frequency filter that overlaps with $S(\omega)$ across a range of frequencies, sometimes extending to $\omega=0$. This short-time behaviour thus contains broad spectral information and can be used to extend the range over which $S(\omega)$ can be determined. In particular, one can choose a drive frequency for which the signal decay is well-matched by the $\chi_2 + \chi_4$ calculation (e.g. $\Omega > 20\:\text{rad/s}$ in figure \ref{fig:GBEErrors}) and extract information about $S(\omega)$ for $\omega < 20\:\text{rad/s}$ from detailed fitting of the short-time behaviour. 

\begin{figure}
\centering
\includegraphics[width = 0.8\linewidth]{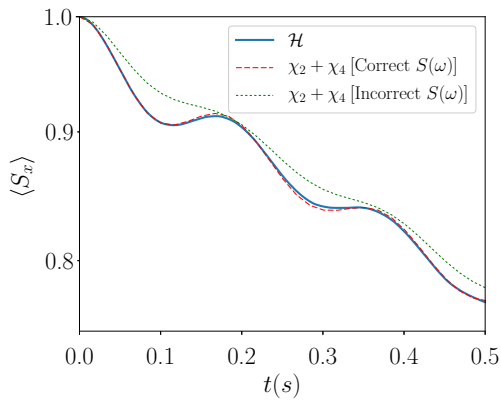}
\caption{\label{fig:wrongS} Comparison of the true signal decay, calculated by unitary time-evolution ($\mathcal{H}$), and decays calculated by the cumulant expansion up to $\chi_4$, when $S(\omega)$ is varied in a region away from the driving amplitude. The correct $S(\omega) = 30\:\textrm{Hz}^2/\omega$ (with a plateau as $\omega \rightarrow 0$) is used to calculate the red curve. For the green curve, all values of $S(\omega)$ for $\omega < 10\:\text{rad/s}$ are reduced by half. The probe frequency, or driving amplitude, is $\Omega = 32\:\textrm{rad/s}$.}
\end{figure}

To illustrate this, figure \ref{fig:wrongS} shows the short-time behaviour of a $\chi_2+\chi_4$ calculated signal decay for two different noise spectra. One spectrum is labelled `correct', while the other represents an error in which $S(\omega)$ at low frequencies has been changed. Here, the error is introduced for $\omega < 10\:\text{rad/s}$, while the drive amplitude $\Omega = 32\:\text{rad/s}$. This shows that the decay is sensitive to variation in $S(\omega)$ at frequencies far below the probing frequency $\Omega$. 

\subsection{Noise spectroscopy protocol}
\label{sec:4.1}
To take advantage of the accuracy of the $\chi_2+\chi_4$ cumulant expansion for improving noise spectroscopy, we propose a gradient ascent protocol based on matching to the experimental signal decay using a single chosen pulse amplitude, $\Omega_P$. An initial estimate of the noise spectrum, $S_0(\omega)$, is obtained from the standard approach of fitting to exponential decays for a group of probe frequencies. Then, by accurately fitting a detailed signal decay at $\Omega_P$ using the cumulant expansion method, particularly in the short-time regime, the full noise spectrum can be determined.

The signal decay given by the cumulant expansion method is 
\begin{align*}
s(t) &= \exp\left( \chi_2 + \chi_4 \right) \\
&= \exp\left( -\frac{1}{\pi} \int_{0}^\infty d\omega S(\omega) \text{Re}(\mathcal{F}_2(\omega,\Omega,t)) \right. \\ & + \left. \frac{1}{2\pi^2} \int_{0}^\infty d\omega_1 \int_{0}^\infty d\omega_2 S(\omega_1) S(\omega_2) \tilde{\mathcal{F}}_4(\omega_1,\omega_2,\Omega,t) \right)
\end{align*}

We discretize the above expression, and define a fitness function as the root-mean square error between the experimentally measured decay, $\left\langle \sigma_x(t_j) \right\rangle$, and the calculated decay $s'(t_j)$ for a given $S'(\omega)$:
\begin{equation}
\Phi = 1 - \sqrt{\frac{1}{n} \sum_{j=1}^n \left(\left\langle \sigma_x(t_j)\right\rangle-s'(t_j)\right)^2}
\end{equation}

We can then calculate the gradient of the fitness function, $\pdiff{\Phi}{S'(\omega_i)}$, for any target frequency $\omega_i$. The gradient is used to update the estimate of $S'(\omega)$ towards a closer matching of the experimental and calculated decays. The full protocol is:

\begin{enumerate}
\item To obtain an initial estimate, $S'_0(\omega)$, use exponential fits of decays over a range of $\Omega$, and matching to $\langle \sigma_x(t) \rangle = \exp(-S'_0(\omega) t / 2)$. 
\item Select a drive amplitude, $\Omega_P$, for detailed matching of the decay curve. $\Omega_P$ should be low enough to display non-exponential features at short times, but not so low that the $\chi_2+\chi_4$ calculation is inaccurate. Otherwise, the following steps will not converge to a high fitness function $\Phi$.
\item Calculate the fitness function, $\Phi_0$, for the decay at $\Omega_P$ using the initial estimate $S'_0(\omega)$.
\item While $\Phi_k < 1-\delta$ for some target threshold $\delta$,
\begin{enumerate}
\item Calculate $\pdiff{\Phi}{S(\omega_i)}$ for the current estimate $S'_k(\omega)$
\item Update the estimate: for all $\omega_i$, $S'_{k+1}(\omega_i) = S'_k(\omega_i) + \epsilon \pdiff{\Phi}{S(\omega_i)}$ for some fixed $\epsilon$
\item Calculate the updated fitness $\Phi_{k+1}$ and increment $k$.
\end{enumerate} 
\end{enumerate}

To improve the speed of calculation/convergence, we can use the knowledge that the simple exponential decay fitting is accurate when the signal decays are smooth exponentials, and only update $S'(\omega_i)$ for $\omega_i$ where that condition is not satisfied. 

\subsection{Demonstration of protocol}
\label{sec:4.2}
\begin{figure}
\includegraphics[width=0.95\linewidth]{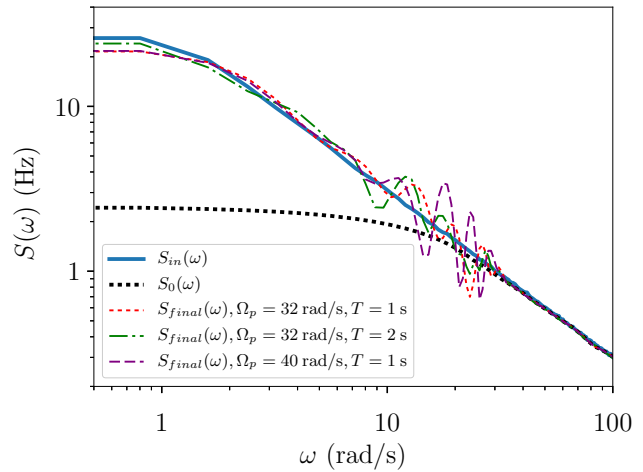}
\caption{\label{fig:resultWpVT} $S(\omega)$ estimates obtained from the cumulant expansion based noise spectroscopy protocol described in the main text. Results shown are for a simulated experiment with $S_{in} = 30\:\text{Hz}^2/\omega$. The final spectrum estimates, $S_{final}(\omega)$, are shown for three different ($\Omega_P$, $T$) conditions, where $T$ is the total pulse duration. The initial estimate $S_0(\omega)$ is determined using the standard method of exponential fits to equation \ref{sigLargeN}. The oscillations that appear in the range of 8-30 rad/s are artifacts (discussed in the text) and can be removed by fitting $S_{final}(\omega)$ to a functional form such as $1/f^k$. }
\end{figure}

The cumulant expansion noise spectroscopy protocol described above was applied to the simulated experiments presented in section 3 corresponding to three different noise spectra. Figure \ref{fig:resultWpVT} shows the result obtained using the simulation with the input noise $S_{input}(\omega) = 30\:\text{Hz}^2/\omega$ (with a plateau as $\omega \rightarrow 0$). The initial estimate, $S_{0}(\omega)$, uses the standard exponential fitting method at 11 pulse amplitudes in the range of 20-125 rad/s.
The final estimate was obtained by detailed fitting to a single decay curve as described above. For comparison, this final step was done with three different choices for the parameters $(\Omega_P, T)$, where $T$ is the total pulse duration. The final noise spectrum estimate $S_{final}(\omega)$ is a much better fit in the low frequency regime to the correct (input) spectrum. 

\begin{figure}
\centering
\includegraphics[width = 0.8\linewidth]{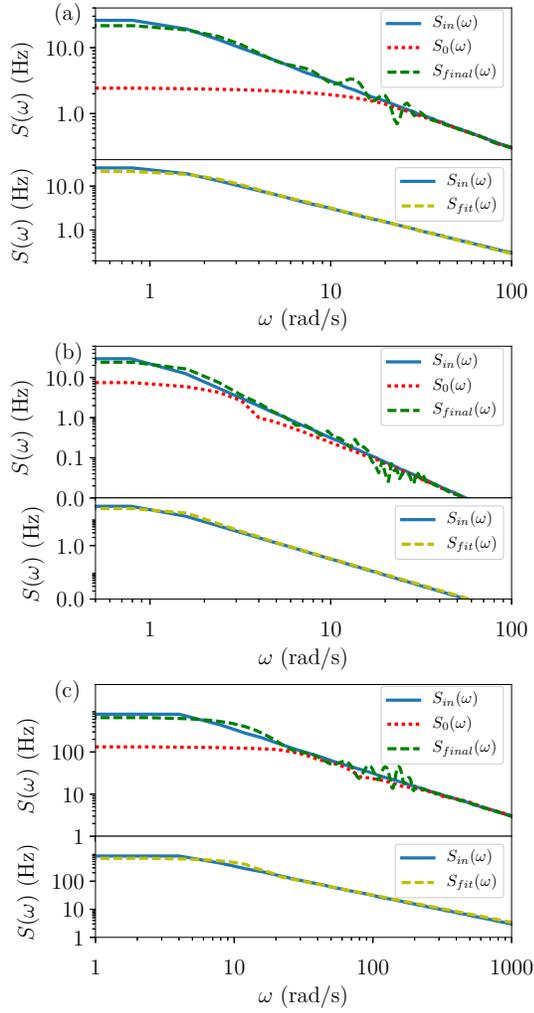}
\caption{\label{fig:result1} The result of applying the cumulant expansion noise spectroscopy protocol to three different noise spectra. The noise spectrum used to simulate each experiment is shown in blue ($S_{in}(\omega)$). The spectrum obtained from the standard exponential fitting ($S_0(\omega)$) is shown in red, and the result of the cumulant expansion protocol ($S_{final}(\omega)$) is shown in green. A fitting of $S_{final}$ to a general power law form $C \cdot \omega^\alpha$ is shown in the lower panels. The input noise spectrum, pulse amplitude $\Omega_P$, and total pulse duration are: (a) $S(\omega) = 30\:\text{Hz}^2/\omega$, $\Omega_P = 35\:\text{rad/s}$, $T = 1\:\text{s}$, (b) $S(\omega) = 30\:\text{Hz}^3/\omega^2$, $\Omega_P = 35\:\text{rad/s}$, $T = 5\:\text{s}$, (c) $S(\omega) = 3000\:\text{Hz}^2/\omega$, $\Omega_P = 640\:\text{rad/s}$, $T = 0.2\:\text{s}$}
\end{figure}

Some artifacts are introduced in the form of oscillations in the intermediate frequency range ($\omega = 8 - 30\:\text{rad/s}$). These artifacts have characteristic periods of order $\sim 2\pi/T$ and are a consequence of remaining error between the $\chi_2 + \chi_4$ decay and the true decay, such as contributions from $\chi_6$ and higher terms. These oscillations are not a consequence of the gradient optimization and we have not found a straightforward way to remove them. However, given that $S_0(\omega)$ is typically a smooth function, and that these oscillations are confined to a certain band of frequencies, smoothing or sliding window averaging can be used to suppress the oscillations in the final estimate. Alternately, they can be fully removed if the noise spectrum can be fit to a certain functional form, such as $1/f^k$. Figure \ref{fig:result1} shows the results of applying the cumulant expansion noise spectroscopy protocol to the same three simulated experiments shown in figure \ref{fig:GBEBreak}. The upper panels show the input, initial, and final $S(\omega)$ determined by fitting the signal decay at pulse amplitude $\Omega_P$ and total time $T$. The lower panels of figure \ref{fig:result1} show the result of fitting $S_{final}(\omega)$ from the upper panel with a general power law $C \cdot \omega^\alpha$, where $C$ and $\alpha$ are free parameters. Note that this power law fit is applied in a frequency range that excludes the plateau region in $S_{final}(\omega)$. To obtain the initial estimate, $S_0(\omega)$, figure \ref{fig:result1}a uses the same experiment set as figure \ref{fig:resultWpVT}, figure \ref{fig:result1}b uses experiments at 15 pulse amplitudes in the range 1-120 rad/s, and figure \ref{fig:result1}c uses 9 pulse amplitudes in the range 10-1250 rad/s.

\section{Conclusion}
\label{sec:5}
In summary, we have treated the problem of spin evolution in the presence of single-axis phase noise during an experiment with CW excitation, with a goal to improve the determination of an arbitrary noise spectral density. By retaining cumulant expansion terms up to fourth order in the system-bath coupling, we can more accurately match coherence decay dynamics that exhibit non-exponential and oscillatory behaviour, and thereby extract more accurate spectral information, especially at low frequencies. We present a two-step protocol: (1) estimate $S(\omega)$ using the standard exponential fitting approach by probing over a set of frequencies (low-resolution signal decay experiments); (2) refine $S(\omega)$ based on fitting a single, high-resolution signal decay using the fourth order cumulant expansion calculation. Since this second step consists of probing at a single frequency, it is efficient in terms of experimental resources. For the cases of $1/f$ and $1/f^2$ noise (with low frequency cutoff plateau), we have shown that this protocol allows for accurate determination of $S(\omega)$ to zero frequency, i.e. the low frequency regime where standard CW and pulsed noise spectroscopy fail. While the examples given were noise spectra of the form $1/f^\alpha$, the theoretical analysis and protocol are applicable to arbitrary spectra, and in future work we plan to test this applicability in simulations and real experiments. In addition, inhomogeneous broadening is typical in physical spin systems, and should also be included. This can be expressed as an additional Hamiltonian component $H(t) = \beta \sigma_z$, where $\beta$ is a static random variable. Thus, inhomogeneous broadening yields a peak in $S(\omega)$ at $\omega = 0$, which should enhance the oscillations in signal decay at short timescales for low probing frequencies. Our protocol should therefore reveal such broadening. Additional work could also extend the cumulant expansion noise spectroscopy protocol to include multi-axis noise and/or higher order cumulants ($\chi_6$) for more general applications.

\begin{acknowledgments}
This work was supported by Natural Sciences and Engineering Research Council (NSERC). D.K.P. was supported by the National Research Foundation of Korea (Grants No. 2015R1A2A2A01006251 and No. 2016R1A5A1008184).
\end{acknowledgments}

\bibliographystyle{unsrt}

\end{document}